# Charge pumping under spin resonance in Si(100) metal-oxide-semiconductor transistors

Masahiro Hori[*] and Yukinori Ono

*Research Institute of Electronics, Shizuoka University, Johoku, Naka-ku, Hamamatsu 432-8011, Japan*

Gate-pulse-induced recombination, known as the charge pumping (CP), is a fundamental carrier recombination process, and has been utilized as a method for analyzing electrical properties of defects (or dangling bonds) at the transistor interfaces, which is now recognized to be well-matured and conventional. Nevertheless, neither the origin (the bonding configuration) of the defects responsible for the CP, nor their detailed recombination sequence has been clarified yet for Si metal-oxide-semiconductor (MOS) interfaces. In order to address these problems, we investigated the CP under spin resonance conditions at temperatures ranging from 27 to 300 K in Si(100) n-type MOS transistors. We obtained evidence that $P_{b0}$ and E' centers, the two major dangling bonds at (and near) the Si(100) interface, participate in the CP recombination process. We also show that the spin-dependent CP process is explained by the formation of electron-electron spin pairs, which in turn reveals that the CP via $P_{b0}$ and E' centers is inherently a two-electron process.

## I. INTRODUCTION

Defects or dangling bonds at the $SiO_2$/Si interfaces [1,2] critically influence the variability and reliability of the metal-oxide-semiconductor (MOS) circuits, and thus have been extensively studied for the past several decades. Even after high-k gate-dielectric materials were introduced [3], defect control of ultrathin $SiO_2$/Si buffer layers is still an essential issue to be studied [4]. Furthermore, rapidly growing Si quantum electronics [5] points out the importance of the $SiO_2$/Si interface states from the viewpoints, e.g., of single-charge [6] and single-spin [7] manipulations, and of control of the valley degree of freedom [8]. In such cases, interface states are not merely "defects", but work as the experimental hosts for the manipulation of the electronic charges, spins, and valleys.

The charge pumping (CP) [9-18], which generates a recombination current by a gate pulse voltage, is a fundamental charge recombination process via interface defects, and has been applied for investigating the interface defects of MOS field-effect transistors (MOSFETs). We expect that the CP could also be a key process for the manipulation of the charges and spins of electrons at the defect sites. The phenomenon was first reported by Brugler and Jespers in 1969 [9], and the basic theoretical model was constructed by Groeseneken et al. in 1984 [10]. Since then, a large number of reports have been issued on this process because it allows us to evaluate various kinds of electrical properties of the interface defects, such as capture cross sections [11,14,16] and density of states profile [10,12,15,18]. However, the origins, or the bonding configurations, of the defects responsible for the CP recombination have not yet been identified. This is because the CP itself does not provide the information for identifying the defects, for which a magnetic resonance technique is called for.

In addition, arguments on the detailed sequence of the CP recombination process have not yet been satisfactorily made. For example, a simple one-by-one electron-hole recombination per defect site is implicitly assumed in conventional defect analysis using the CP (see, e.g., the potential diagram shown in Fig. 1(b)). However, as is well accepted in the literature, the dangling bonds of the host material, Si in the present case, have three states; positively charged no-electron, neutral one-electron, and negatively charged two-electron states [7,19,20], and consequently, there are two types of energy levels, corresponding to the transitions +/0 and 0/– [20,21]. The CP should reflect this amphoteric nature. Actually, the possibility of the multiple-electrons CP via one defect site has been pointed out in the early stage of the CP research [13], and a recent article reported that one defect site at $SiO_2$/Si interfaces can convey two electrons at most in one cycle of the CP sequence [17].

In order to address these unsolved problems, we performed and analyzed the CP under electron-spin resonance (ESR) conditions, which we here refer to as CP EDMR (electrically-detected magnetic resonance). Contrary to the conventional ESR technique [22], where the absorption of the microwave power is measured, the EDMR [23] measures the current, and thus enables us to make a selective and sensitive detection of the spins located in the area where the current is limited. The CP EDMR literally measures the CP recombination current using the EDMR technique and is expected to give us information about the dangling bonds responsible for the CP process.

The CP EDMR was first applied in 1998 to a Si MOSFET [24] (and then to SiC MOSFETs [25,26]) at room temperature. However, no decisive conclusion about the origin and detailed mechanism of the CP process has been made because of the poor signal to noise ratio [24]. We therefore report here the CP EDMR on Si (100) MOSFETs in the temperature range between 27 and 300 K. Thanks to the enhanced signal intensity at low temperatures and also to our low-noise measurement system [27], we obtained signals strong enough for detailed analysis. We show that two major dangling bonds observed so far using conventional ESR on (100) SiO$_2$/Si interfaces, P$_{b0}$ [28,29] and E' [29,30], participate in the CP recombination. We also show that the present CP EDMR data follows the electron-electron spin pair model [31], which in turn supports the idea that the CP via these defects is a two-electron process.

---

*hori.masahiro@shizuoka.ac.jp

## II. RESULTS

### A. Measurement setup and device structure

Figure 1(a) shows the setup for the (conventional) CP measurements. In the CP, a pulse voltage with a large amplitude is repeatedly applied to the MOS gate so that the electron inversion and hole accumulation layers are alternatively formed. The CP sequence in one cycle of the pulse is shown in Fig. 1(b); conduction-band electrons are captured by interface defect states in the rise step of the pulse, and the trapped electrons subsequently recombine with valence-band holes in the fall step. The chain of the pulse causes the electron-hole recombination current to flow between the source/drain and the substrate via the interface defects. This is called CP current, given by $I_{cp} = efAN_{it}$, where $e$, $f$, $A$, and $N_{it}$ are the elementary charge, the gate pulse frequency, the channel area, and the area density of the interface states, respectively.

In this study, n-type MOSFETs fabricated on a Si (100) substrate were used. The channel length/width and gate oxide thickness are respectively 50/500 μm and 30 nm, and the substrate doping (boron) concentration is of the order of $10^{15}$ cm$^{-3}$. The gate oxide thickness of 30 nm was chosen because, with this oxide thickness, the charge pumping measurements can be performed without disturbance by the gate leakage current with a moderate level of the gate-voltage swing. The gate oxide was formed in a dry oxygen ambient at 950 °C for 50 minutes, and the fabrication process was finalized with the forming gas (N$_2$:H$_2$ = 2:1) treatment at 450 °C for 30 minutes. The gate was made of phosphorus-doped poly-Si.

Here, we intentionally increased the interface defect density by applying the Fowler–Nordheim (FN) stress [32]. This is because the defect state density of the initial (fresh) interfaces was too low (on the order of $10^9$ cm$^{-2}$ at room temperature) to obtain signals sufficiently strong for the analysis of superimposed peaks (see below), and also because we can make certain that the near-interface states, which are observable only in stressed samples [33], participate in the CP process. We obtained samples with $N_{it}$ at around $10^{11}$ cm$^{-2}$. Details of the FN stress application and the resultant CP EDMR data can be found in the APPENDIX.

For the CP EDMR, a MOSFET was inserted into the TE$_{011}$ cylindrical cavity of the X-band ESR system (Fig. 1(c)), and was irradiated by microwave with the power of 100 mW. The $V_{base}$ was kept at the constant value at which $I_{cp}$ reaches its maximum value $I_{cp,max}$, and the output signal (the CP current) was recorded by using a lock-in amplifier with the magnetic-field modulation. The magnetic field was modulated with the frequency and amplitude of 100 Hz and 2.0 G, respectively. The sample was cooled down with a closed-cycle He cryo-cooler, and the measurements were done at temperatures ranging from 27 to 300 K. Note that the CP and the CP EDMR measurements were difficult to perform at temperatures below 27 K for the present MOSFETs with the low doping density of the substrate because the substrate became insulating due to the carrier freezeout.

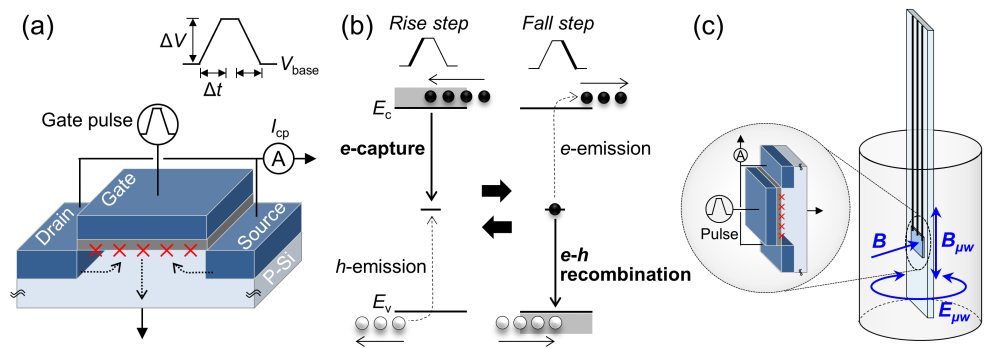

FIG. 1. Setup for the CP and CP EDMR. (a) Setup for the charge pumping (CP). Dashed arrows show the electron flow between the source/drain and the substrate via the interface defects (red cross marks). The parameters of the gate pulse are defined in the inset of the figure, where Δ$V$, Δ$t$, and $V_{base}$ are the amplitude, the rise/fall time and the base voltage of the pulse, respectively. The duty cycle of the pulse was fixed at 50% for all measurements. (b) CP sequence. Electrons (closed circles) are captured by the defect states in the rise step of the pulse (left), and then the trapped electrons recombine with valence-band holes (open circles) in the fall step (right). The electron (hole) emission, which is the competing process against the recombination (electron capture) process, is indicated by the dotted arrow. $E_c$ and $E_v$ are the edge of the conduction and valence band, respectively. Note that the diagram is drawn by assuming one-electron capture/recombination model, implicitly assumed in the literature. On the other hand, we here propose a two-electron CP process, see Fig. 7(b). (c) Setup for CP EDMR. A MOSFET is mounted on the sample holder and inserted into the cylindrical cavity of the ESR system. The arrow $B$ shows the static magnetic field while the arrows $B_{\mu w}$ and $E_{\mu w}$ show the microwave magnetic and electric fields, respectively.

## B. CP and CP EDMR measurements

We first show the results for the conventional CP. Figure 2(a) shows the CP current as a function of base voltage $V_{base}$ of the pulse at the temperatures ranging from 27 to 300 K. The definition of the $V_{base}$ is shown in Fig. 1(a). Figure 2(b) shows $I_{cp,max}$ (left axis) and $N_{it}$ estimated from the $I_{cp,max}$ (right axis). One can see that $I_{cp,max}$ and hence $N_{it}$ decrease with the temperature.

For the CP, the temperature is an important parameter. Energy levels of the interface defects are widely distributed in the bandgap, and change in the temperature enables us to control the range of the energy levels of the defects that can participate in the CP process. The above results indicate that the energy range for the CP becomes narrower as the temperature increases [12] (See the diagram shown in the inset of Fig. 2(b).)

As illustrated in Fig. 1(b), right (fall step), when the gate voltage is switched from a positive to a negative value, conduction-band electrons escape back to the source/drain, leaving the trapped electrons behind, and holes are subsequently accumulated at the interface, resulting in the recombination of the trapped electrons with the holes. During this CP recombination process, emission of the trapped electrons to the conduction band could occur (dotted arrow in Fig. 1(b), right) and work as a competing process against the recombination. If the emission took place, the trapped electron escapes back to the source/drain, and does not contribute to the CP current. Because the emission is thermally activated, a shallower defect level has a higher emission rate. Therefore, there is a boundary of the energy level above which the emission rate becomes higher than the recombination rate, and, as the temperature increases, this energy boundary shifts downward due to the enhanced emission rate. A similar competing process, the hole emission, takes place for the electron capture process shown in Fig. 1(b), left, which results in the shift of the lower bound of the boundary energy upward with the temperature. As a result, the range of the defect energy levels for the CP becomes narrower with the temperature. Following ref. [12], the energy range is estimated to be $E_i \pm 0.25$ eV for 300 K and $E_i \pm 0.55$ eV for 27 K, where $E_i$ is the Fermi level of intrinsic Si. As it will be explained below, the CP EDMR signals are expected to come from the defects whose energy level is located around the upper energy boundary, and from the above estimation, one can understand that the present temperature variation (27 – 300 K) scans the wide range of the energy levels of the defect states.

Figure 3(a) shows the output spectrum of the CP EDMR measured at 27 K as a function of the magnetic field $B$, which was applied perpendicular to the (100) interface, or parallel to the [100] direction ($\boldsymbol{B}\|[100]$, see the Fig. 1(c)). The data were taken at the constant $V_{base}$ (= –5 V) at which the $I_{cp}$ reaches $I_{cp,max}$ (see inset of Fig. 3(a)). By integrating this differentiated output spectrum with respect to $B$, we can obtain the actual change $\Delta I_{cp}$ in the $I_{cp}$ caused by the resonance, which is shown in Fig. 3(b). We here point out that $\Delta I_{cp}$ is positive. This means that the CP current increases when the resonance occurs.

As shown in Fig. 1(b), the electron capture (in the rise step of the pulse) and subsequent electron-hole recombination (in the fall step) has a competing process, the hole and electron emission, respectively, and the rates of these four processes determine the energy range of the defects that can contribute to the CP current. Therefore, the positive $\Delta I_{cp}$ strongly suggests that the energy window for the CP becomes wider because of at least one of these rates varying, e.g., the electron-hole recombination rate being enhanced, due to the resonance. In other words, electrons (or holes) that otherwise were supposed to be emitted participate in the CP recombination by the resonance. This means that the CP EDMR signals come only from the defects whose energy is around the energy boundaries while the CP current is dominated by all the defects within the energy range. As it will be shown later, we ascribed the positive $\Delta I_{cp}$ to the enhancement of the electron-hole recombination rate due to the resonance, which shifts the upper bound upward. Note that $\Delta I_{cp}$ is on the order of pA while $I_{cp,max}$ is on the order of 100 nA ($\Delta I_{cp}/I_{cp,max}$ is on the order of $10^{-5}$), indicating that the recombination rate enhancement is quite small. However, this intensity is strong enough to make a detailed discussion. Discussion about the mechanism of the positive $\Delta I_{cp}$ will be made later, and we will first discuss the origin or the bonding configuration of the defects responsible for the CP EDMR signal.

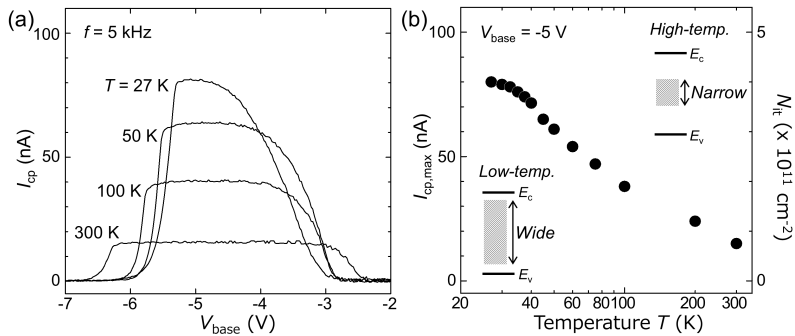

FIG. 2. CP characteristics. (a) Temperature dependence of the CP current $I_{cp}$ as a function of the base voltage $V_{base}$ of the pulse. (b) The maximum value $I_{cp,max}$ and area density $N_{it}$ of the interface defects estimated from the $I_{cp,max}$. Inset shows the range of the energy levels of the defects that can contribute to the CP current.

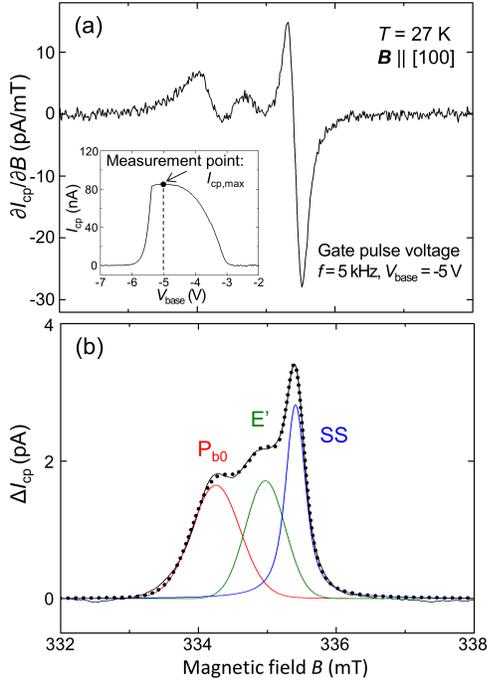

FIG. 3. CP EDMR characteristics. (a) Output (differential) spectrum measured at 27 K with the magnetic field $B$ parallel to the [100] direction ($\mathbf{B}\|[100]$). Parameters for the CP gate pulse are $f = 5$ kHz, $\Delta V = 4$ V, and $\Delta t = 30$ μs. $V_{base}$ is fixed at −5 V, at which the $I_{cp}$ is maximum (= 80 nA) as shown by the arrow in the inset. (b) Integrated spectrum. Thin solid black curve is the experimental data. This curve is deconvoluted into three Voigt functions for $P_{b0}$ (red line), E'(green), and SS (blue). Their sum is shown by the dotted black curve.

As one can see in Fig. 3(b), the signal is composed of several components. We deconvoluted it into Voigt functions [34,35], and found that three Voigt functions (three peaks) gave us the best fit (dotted black curve). The resultant $g$-values of the peaks agree with those of $P_{b0}$ (red line, $g$ = 2.006) [28,29], E' (green line, $g$ = 2.002) centers [29,30], and weakly-localized electrons at shallow states around the conduction-band edge (blue line, $g$ = 1.999) [36,37]. (We hereafter refer to these electrons as weakly-localized shallow-state electrons, or simply as SS electrons.) Here, $P_{b0}$ is the Si dangling bond right at the (100) interface with three Si backbonds [28,29], while the E' is the Si dangling bond at an oxygen vacancy in the $SiO_2$ film [29,30].

Note that we did not use the Dysonian profile, which is a typical feature for the spins of mobile carriers [38], as a fitting curve, because no mobile electrons are expected to be left in the conduction band when the recombination takes place in the CP process. Also note that, although the E' centers are located in the oxide, they can be detected in the CP measurements if the gate-pulse frequency is low [39], and the present frequency, $f$ = 5 kHz, is low enough for the E' center detection.

Figures 4(a) and 4(b) show the output and the integrated spectra for five rotation angles of the sample with respect to the magnetic field, ranging from $\varphi$ = 0° ($\mathbf{B}\|[100]$) to 90° ($\mathbf{B}\|[011]$). One can see that spectra in the region below 335 mT (shaded area in Fig. 4(a)) have a considerably large rotation-angle dependence. This is consistent with the fact that the bonding directions of the $P_{b0}$ centers (red) are anisotropic (while the E' and the SS are isotropic due to the randomly-oriented orbitals and due to an $s$-like orbital, respectively).

According to the previous ESR data and theory [28,29], the single peak of the $P_{b0}$ centers at $\varphi$ = 0° splits into three with the intensity ratio of 1:1:1 by the rotation (see red dotted curves in Fig. 4(c)). We searched the best fitting of the intensity curves, using the $g$-value and width of each peak as variables, while the peak heights of the E' and SS are unchanged and those of the three (split) peaks of the $P_{b0}$ are kept at 1/3 of that for $\varphi$ = 0°. The results of the fitting are indicated by dotted curves in Fig. 4(b) and the resultant $g$-values for five rotation angles are plotted in Fig. 4(c). One can see that the fitting (dotted curve) reproduces the experimental data (thin black line) for all the rotation angles, which supports the validity of the assignment.

We mention that we could not obtain a reasonable fitting for all the rotation angles ($\varphi$ = 0° ($\mathbf{B}\|[100]$) to 90° ($\mathbf{B}\|[011]$)) using the theoretically predicted $g$-tensor for the $P_{b1}$ center [40], another type of the dangling bond at the (100) $SiO_2$/Si interface. For example, for $\varphi$ = 0° ($\mathbf{B}\|[100]$), $P_{b1}$ center has two peaks with the $g$-values of 2.0058 and 2.0041 [40], but it was difficult to reproduce the centered hump (which we ascribed to the E' center) with these $g$-values.

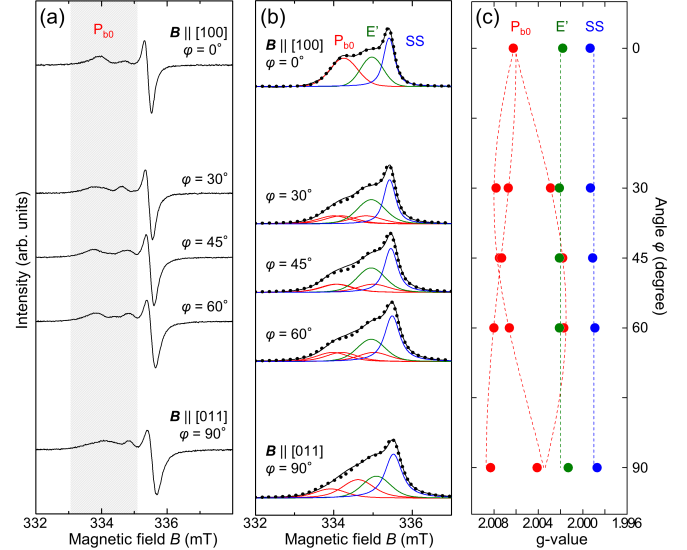

FIG. 4. Rotation-angle dependence of CP EDMR signals. (a),(b) Output and integrated spectra for five rotation angles ranging from $\varphi$ = 0° ($\mathbf{B}\|[100]$) to 90° ($\mathbf{B}\|[011]$). Parameters for the CP gate pulse are the same as those for Fig. 3 ($f$ = 5 kHz, $\Delta V = 4$ V, $\Delta t = 30$ μs, and $V_{base}$ = −5 V). In (b), thin solid black curves are the experimental data, and they are deconvoluted into Voigt functions for $P_{b0}$ (red lines), E'(green line), and SS (blue line). Their sums are shown by the dotted black curves. (c) Rotation-angle dependence of the $g$-values. Dots are the $g$-values extracted from the experimental data, while the curves are the theoretically predicted ones. Red, green and blue dots and curves are, respectively, for $P_{b0}$, E', and SS.

Figure 5(a) shows the integrated spectra (black line) and the decomposed peaks (red, green, and blue lines) for five CP-pulse frequencies $f$'s ranging from 3 to 5 kHz with $B\|[100]$ ($\varphi = 0°$). The peak heights, $\Delta I_{cp}(P_{b0})$, $\Delta I_{cp}(E')$, and $\Delta I_{cp}(SS)$ of the peaks are plotted as a function of $f$ in Fig. 5(b). The $I_{cp,max}$ is also plotted in the figure. One can see that they are proportional to $f$, which demonstrates that $P_{b0}$, E' and SS participate in the CP process. We mention, however, that the SS is not the recombination center (but the partner of the $P_{b0}$ and E' centers for constituting a spin pair), as we will explain later.

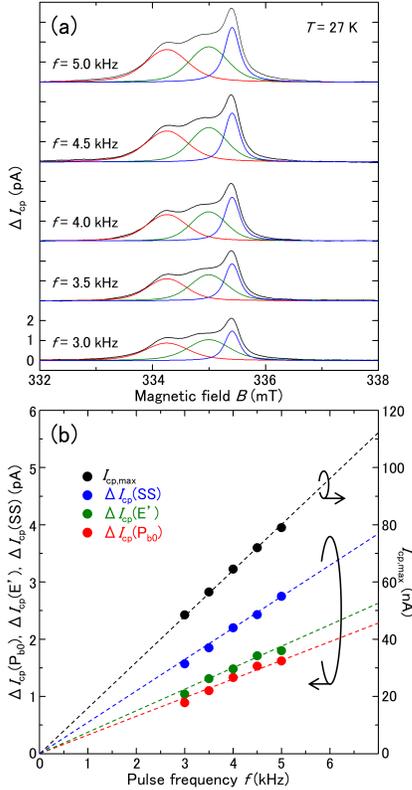

FIG. 5. Pulse-frequency dependence of the CP EDMR signals. (a) Integrated spectra (black lines) for gate pulse $f$ ranging from 3 to 5 kHz with $B\|[100]$, each of which is deconvoluted into three Voigt functions for $P_{b0}$ (red), E' (green), and SS (blue). Except for $f$, the pulse parameters are the same as those of the data shown in Fig. 3 ($\Delta V = 4$ V, $\Delta t = 30$ μs, and $V_{base} = -5$ V). (b) Peak height of the Voigt function for $P_{b0}$ (red), E' (green), and SS (blue) as a function of $f$. $I_{cp,max}$ (black circles) is also plotted.

We next show the temperature dependence. Figure 6(a) shows the CP EDMR spectra for temperatures $T = 27, 60,$ and 300 K for $B\|[100]$ ($\varphi = 0°$). One can see that $P_{b0}$ and E' centers are detected for all the temperatures with comparable intensities to each other. As we have touched on, and will explain in detail in the Discussion section, the CP EDMR signals are expected to come from the defects whose energy is around the upper bound of the energy range for the CP. Therefore, the data indicates that our assignment holds true for the wide range of the defect-state energy. We estimated the energy boundary by using Eq. 19 of Ref. [12], and by assuming the Si bandgap to be 1.17 eV at 27 and 60 K, and 1.12 eV at 300 K [41]. Results are shown in the inset of Fig. 6(a).

We should mention that, according to Ref. [42], the energy levels of the $P_{b1}$ centers are concentrated around the midgap, and they are not on the energy boundaries for the present temperature range. This will be the reason why we did not observe the $P_{b1}$ centers in the CP EDMR signals.

We show the temperature dependence of $\Delta I_{cp}(P_{b0})$ and $\Delta I_{cp}(E')$ in Fig. 6(b), and of $\Delta I_{cp}(SS)$ in Fig. 6(c). We also show the ratio $R$ of $\Delta I_{cp}(SS)$ to $\Delta I_{cp}(P_{b0}) + \Delta I_{cp}(E')$ in the inset of Fig. 6(c). We found that neither $\Delta I_{cp}(P_{b0})$, $\Delta I_{cp}(E')$, nor $\Delta I_{cp}(SS)$ is simply proportional to $I_{cp}$ or to $dI_{cp}/dT$ (which reflects the density of states of the defect levels). Rather, the temperature dependence of the $\Delta I_{cp}$'s is strong and complicated, and the temperature range can be classified into three regions, $T \lesssim 40$ K, $40$ K $\lesssim T \lesssim 80$ K, and $T \gtrsim 80$ K, depending on the behavior of the SS signals. The signal from the SS sharply decreases with increasing $T$ in the range $40$ K $\lesssim T \lesssim 80$ K, and are undetected for $T \gtrsim 80$ K. From this temperature dependence, we can make a detailed argument on the spin-dependent process, which will be shown below.

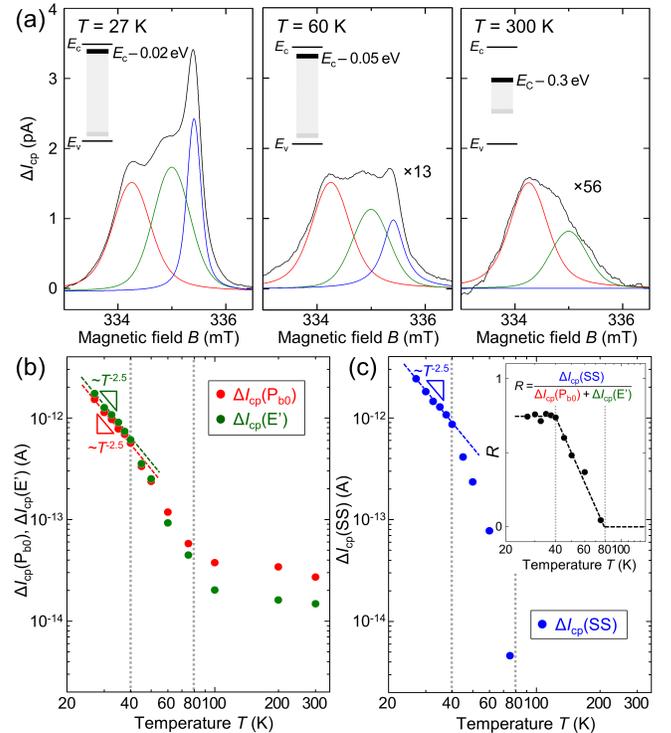

FIG. 6. Temperature dependence of the CP EDMR signals. (a) Integrated spectra $\Delta I_{cp}$ for temperatures $T = 27, 60$ and 300 K with $B\|[100]$. Each spectrum (black line) is deconvoluted into three Voigt functions, each of which is assigned to $P_{b0}$(red), E'(green), and SS(blue). Data for $T = 60$ K and 300 K are respectively expanded to the vertical direction by 13 and 56 times for clarity. CP pulse parameters are the same as those of the data shown in Fig. 3 ($f = 5$ kHz, $\Delta V = 4$ V, $\Delta t = 30$ μs, and $V_{base} = -5$ V). In the inset of each graph, the shaded area and bold line indicate the energy range for the CP and the energy level for the CP EDMR, respectively. (b) $\Delta I_{cp}$ as a function of $T$ for $P_{b0}$ (red) and E' (green) centers. (c) $\Delta I_{cp}$

as a function of $T$ for SS. Inset in (c) shows the ratio $R$ of $\Delta I_{cp}$(SS) to the sum of $\Delta I_{cp}$($P_{b0}$) and $\Delta I_{cp}$(E') as a function of $T$. In (b) and (c), two vertical dotted lines define three temperature regions ($T \lesssim 40$ K, $40$ K $\lesssim T \lesssim 80$ K, and $T \gtrsim 80$ K).

## III. DISCUSSION

### A. Analysis of the CP EDMR data

In conventional ESR [22], the signal comes from the microwave absorption and subsequent energy dissipation at "individual" spins. In the EDMR, on the other hand, it comes from the modulation of the current due to the interplay of "two spins" [23,31,43-54]. The key concept of the spin-dependent process in the EDMR is the spin pair, which constitutes either a singlet or a triplet state. Note that, in the literature of EDMR, the terms "singlet" and "triplet" indicate the configuration (alignment) of two spins forming the pair, and does not necessarily mean the quantum eigenstates of the two spins [47,50]. In other words, we term the spin pair "singlet" ("triplet") if the two spins are antiparallel (parallel), irrespective of the strength of the spin-spin interaction. The entire recombination process would become spin dependent if the singlet and triplet states have a different transition rate in either electron capture or electron-hole recombination step, and magnetic resonance signals are then observable because the spin resonance changes the population of these states.

There are two kinds of spin pairs reported, the electron-electron [31,43,48,50-54] and electron-hole [23,44-46,49] pairs. They were introduced to explain the spin-dependent photo-conductivity in amorphous, micro-crystalline and crystalline Si. As is well known, the dangling bonds of the host material, Si in the present case, have three states, positively charged no-electron, neutral one-electron, and negatively charged two-electron states [7,19,20]. Therefore, with one electron in the defect site, the second electron can be captured. During such electron capture process, the electron-electron spin pair can be formed. On the other hand, when the electron-hole recombination takes place, the defect electron could strongly couple with a valence-band hole, which constitutes an electron-hole spin pair.

The present CP EDMR data are explained by applying the electron-electron pair model to $P_{b0}$ and E' centers. The situation is illustrated in Fig. 7(a). The second electron is not directly trapped at the ground state of the defect ($P_{b0}$ or E'), which is the singlet state, but is first trapped to some intermediate excited states, forming the intermediate pairs, which can be in a singlet or in a triplet state [31,50-53]. The intermediate pair can be two spins in the doubly-occupied excited states of the defects or the ones composed of a singly occupied defect state and a closely located paramagnetic state such as the SS [48,50,51]. We here refer to these pairs as "on-site" and "off-site" pairs, and, as will be explained below, these pairs predominate in the high- and low-temperature regimes, respectively. (See Fig. 7(a).) We should mention that the off-site pair has already been observed in the measurement of the photo-excited current of a Si surface with a native oxide. In that case, the pair partners were the electron spins of phosphorus donors doped in the Si substrate [53,54]. This means that a deep state and a shallow state can constitute the off-site pair.

Once the second electron is trapped at a triplet intermediate state, it is prohibited to make a transition to the ground (singlet) state due to the spin conservation law [31,51,52]. Since the ground state has a higher recombination rate than those of the intermediate states (because it is the deepest level), this spin selection rule makes the entire recombination process spin dependent. That is, the spin resonance induces the triplet-to-singlet transition at the intermediate pairs, which permits the pairs to be relaxed down to the ground singlet states, enhancing the net recombination rate.

In the present CP EDMR case, this enhancement should give a significant impact on the electrons trapped at the defects whose (ground state) energy is around the boundary for the CP. The resonance permits these electrons, which would otherwise have emitted back to the source/drain, to join the recombination activity, leading to the increase in the CP current, or to a positive $\Delta I_{cp}$. The electrons trapped at defects with deeper ground states would not be affected by the resonance because the emission rate is low and thus they can recombine with a high probability without the help of the recombination-rate enhancement. This leads us to the idea that the CP EDMR dominantly detects defect states whose ground-state energy is around the upper boundary.

We propose that the off-site and on-site pairs predominate at low- ($T \lesssim 40$ K) and high- ($T \gtrsim 80$ K) temperature regimes, respectively. In particular, in the low-temperature regime, an SS electron constitutes the off-site spin pair either with a $P_{b0}$- or E'-center electron. In other words, the SS electrons are the pair partners of the $P_{b0}$ or E' electrons, which means that the SSs themselves are not the recombination centers.

There are several reasons for this assignment. One is the fact that the $R$ (= $\Delta I_{cp}$(SS)/($\Delta I_{cp}$($P_{b0}$) + $\Delta I_{cp}$(E'))) is nearly constant for $T \lesssim 40$ K, and its value ($\simeq 0.75$) is on the order of unity (Fig. 6(c) inset). It is noteworthy that this result holds true for a different sample with a different intensity ratio between $P_{b0}$ and E' centers. (We can control the intensity ratio between the two centers by controlling the FN stress conditions. See the APPENDIX.) In addition, the signal intensity of the SS was found to be negligibly small for unstressed samples where the signal intensities of $P_{b0}$ and E' centers were also negligibly small (data not shown). In other words, the SS signals are detected only when the signals from the $P_{b0}$ and E' centers are present, and their intensity is comparable to the sum of the intensities of the $P_{b0}$ and E' centers.

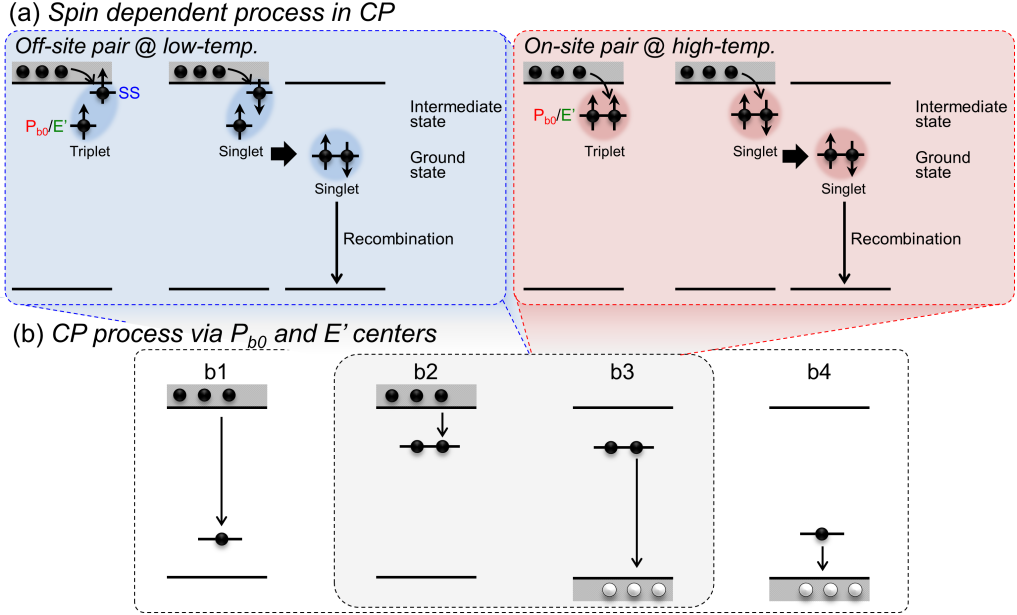

FIG. 7. Proposed model of the CP process. (a) Two types of electron-electron spin pairs, off-site (left) and on-site pairs (right), are dominantly formed at low- and high-temperature, respectively. The triplet/singlet intermediate pairs are forbidden/allowed to make a transition to the ground states. (b) CP process via $P_{b0}$ and E' centers. Steps b2 and b3 are relevant to the present CP-EDMR signals.

Another reason for the assignment is the temperature dependence of the three signals. As shown in Figs. 6(b) and 6(c), the defect centers ($P_{b0}$ and E') and the SS have nearly the same temperature dependence for $T \lesssim 40$ K; the $\Delta I_{cp}$'s change as $T^{-\alpha}$ with $\alpha \simeq 2.5$. Moreover, $\Delta I_{cp}(P_{b0})$ and $\Delta I_{cp}(E')$ deviate from this simple form at $T \simeq 40$ K, which coincides with the temperature at which $\Delta I_{cp}(SS)$ starts to sharply decrease. This means that the signal intensity of the SS has a strong correlation with those of the $P_{b0}$ and E' centers. All these data support the idea that the SS-electrons are the pair partners of the $P_{b0}$ and E' electrons.

Then, the temperature dependence ($T^{-\alpha}$ with $\alpha \simeq 2.5$) strongly suggests that the thermal polarization of the spin pairs [23,43] predominate in the intensity of the CP EDMR signals. With a static magnetic field $B$, the Zeeman splitting of the energy $E_0 = h\nu_0 (= g\mu_B B)$ results in the polarization of a spin ensemble, where $h$, $\nu_0$, $g$, $\mu_B$ are the Planck constant, Zeeman frequency, $g$ factor and Bohr magneton, respectively. Under the thermal equilibrium conditions, the polarization $p$ is given by $p \simeq h\nu_0/2kT$ for $h\nu_0 \ll kT$, where $k$ is the Boltzmann constant [23]. For an ensemble of spin pairs, the polarization $p_{pair}$ (= $p_t - p_s$) is given by $p_{pair} \simeq (h\nu_1/2kT)(h\nu_2/2kT)$, where $p_t$ and $p_s$ are the probability that a pair is found in a triplet or in a singlet state, respectively, and $\nu_1$ and $\nu_2$ are the Zeeman frequencies of the two spins in the pair [46]. One can see that we find a triplet pair more probable than the singlet pair, and the $p_{pair}$ in the thermal equilibrium is scaled as $T^{-2}$. Under the resonance conditions of either spin in the pair, the corresponding $p$ is forced to be zero because of the Rabi oscillation, which reduces $p_t$ (increases $p_s$). This results in the enhancement of the recombination rate, thus of the CP current. This is why we observe EDMR signals for both resonance conditions of the defects ($P_{b0}$ or E') and the SS. Since the intensity is expected to be proportional to $\Delta p_{pair}$ = $p_{pair}$ (thermal equilibrium) − $p_{pair}$ (resonance), the $\Delta I_{cp}$ would also follow the $T^{-2}$ law. Note that the observed exponent $\alpha$ ($\simeq 2.5$) was slightly larger than 2. This may be because of the spin-lattice relaxation [23]. The spin-lattice relaxation results in the transition from the triplet to the singlet states even under the non-resonance conditions. This "leakage" will be suppressed efficiently at a lower temperature due to the longer relaxation time.

In the temperature range, $40$ K $\lesssim T \lesssim 80$ K, the signal from the SS sharply decreases with increasing $T$. This will be because, due to the shallow nature of the states, the lifetime of the state becomes shorter than the time constant of the capture process of the SS-electrons to the defects ($P_{b0}$ or E' sites). Accordingly, the temperature dependence of the $\Delta I_{cp}$ for $P_{b0}$ or E' deviates from the $T^{-2.5}$ slope.

In the temperature range $T \gtrsim 80$ K, the signal intensity of the SS is less than the detectable limit ($\Delta I_{cp} \lesssim 5$ fA). This indicates that the off-site pairs are hardly formed. Then, the on-site pairs become apparent, which can be formed by the direct capture of an electron from the extended conduction-band states (see Fig. 7(a)). One can see in Fig. 6(b) that the temperature dependence of $\Delta I_{cp}(P_{b0})$ and $\Delta I_{cp}(E')$ becomes weak for $T \gtrsim 80$ K; it is nearly independent of $T$. This suggests that the Kaplan-Solomon-Mott (KSM) mechanism [44] is effective in this temperature range. The key idea of the KSM mechanism is the exclusivity [31,46,50]. That is, once two electrons form a spin pair, and if the pair was a triplet, no recombination process proceeds unless the pair is dissociated into two independent spins for a new pair to be formed. In such a case, the temperature dependence is lost.

## B. Model of the CP process

We now propose in Fig. 7(b) the entire CP process via $P_{b0}$ and E' centers. Based on the above analysis, we conclude that the CP via $P_{b0}$ and E' centers is necessarily a two-electron process; the first and second electrons are sequentially captured to a defect site (b1), (b2), and the captured electrons subsequently recombine with the valence-band holes one by one (b3), (b4).

The capture process of the first electron (step b1) cannot be spin-dependent. This is because no electrons are trapped at the initial state, and thus spin pair model cannot be applied. (Note that Zeeman splitting of the one-electron state at the defect sites would not affect the capture rate. This is because the Zeeman energy is on the order of several tens μeV, and this value is much smaller than the ionization energy even for the shallow defects. The capture rate of the first electron would thus have no significant difference between the spin up and down states.)

Recombination of the remaining (second) electron (step b4) is not spin-dependent either. At this step, the electron-hole spin pair could be formed and thus the electron-hole recombination rate could be spin-dependent. However, no signals from the valence-band holes ($g \sim 2.065$-$2.08$, $2.0095$-$2.0105$) [45,49] were detected in the present CP EDMR measurements (e.g., shown in Fig. 3). This indicates that electron-hole spin pair does not account for the present data.

The above argument leads us to the conclusion that one-electron steps (steps b1 and b4) are not relevant to the present CP EDMR signals. In other words, the simple one-electron capture/recombination model shown in Fig. 1(b) cannot describe the CP recombination via $P_{b0}$ and E' centers.

As we have explained, the CP EDMR signals were positive. This indicates that the resonance increases the CP recombination rate by forcing the transition of the triplet states to the singlet states. Increase in the CP current ($\Delta I_{cp}/I_{cp,max}$) was on the order of $10^{-7}$ – $10^{-5}$, depending on the temperature. Such small values indicate that the recombination-rate enhancement is quite small. However, this change, though it is quite small, is a strong implication that the spin configuration inherently influences the CP recombination even with the conventional setup without the magnetic field because the triplet pairs can still be formed under such conditions. This spin-dependent property therefore will be critically important when discussing the accurate transfer of elementary charges [6] by the CP recombination.

## IV. SUMMARY

We performed CP EDMR on the (100) $SiO_2$/Si interfaces of the Si MOSFETs at temperatures ranging from 27 to 300 K. We showed that $P_{b0}$ and E' centers are dominantly responsible for the CP current. In addition, we showed that the spin-dependent process during the CP sequence is the capturing of the second electron to the defects and subsequent recombination, from which we concluded that the CP via $P_{b0}$ and E' centers is a two-electron process.

## ACKNOWLEDGMENTS

The authors thank A. Fujiwara for his technical support, and M. Tabe, T. Tsuchiya, D. Moraru, K. Ando, G. Ferrari, T. Umeda, and Y. Yonamoto for their valuable comments. This work was partially supported by JSPS KAKENHI (JP16H02339, JP16H06087, and JP17H06211), JST CREST (JPMJCR1774), JSPS Bilateral Joint Project between JSPS and CNR in Italy, and the Cooperative Research Project of Research Center for Biomedical Engineering and of Research Institute of Electronics, Shizuoka University.

## APPENDIX: CP EDMR SPECTRA FOR FOWLER-NORDHEIM STRESSED MOSFETS

Fowler-Nordheim (FN) tunneling is the electron tunneling through a triangular potential barrier and is observed when a high electric field is applied to a $SiO_2$ dielectric film [55,56]. The FN tunneling causes the generation of point defects at and near the Si/$SiO_2$ interface [57-59], and the defects generated by the FN tunneling are called FN-stress-induced defects [57]. The mechanism of the defect generation has been intensively studied from the view point of the hot-carrier issue [58], and now is believed to be relevant to the electron-hole pair creation by the high-energy tunneling electrons and subsequent injection of hot holes (and electrons) into the $SiO_2$ film and its interface [59].

In this APPENDIX, we compare the CP EDMR spectra of MOSFETs with different FN-stress conditions. We show the dependence of the CP EDMR signals on (i) injected charges and on (ii) stress-bias polarity.

(i) *Dependence on injected charges*

The results are summarized in Figs. 8 and 9. Figure 8 shows the area defect density $N_{it}$ estimated from the CP measurements as a function of the injected charges at room temperature. For the FN stress, a negatively-high voltage (fixed at $V_g = -30$ V) was applied to the gate, which corresponds to the electric field of $-10$ MV/cm applied to the oxide. One can see that $N_{it}$ is saturated at a value slightly larger than $1 \times 10^{11}$ cm$^{-2}$. The data displayed in the main text was obtained for a sample whose $N_{it}$ is close to this saturation value, which is denoted by Tr. A in Fig. 8.

Figure 9 shows the CP EDMR spectra at 27 K for the transistor (Tr. A in Fig. 8) used in the main text and the one (Tr. B) with a smaller $N_{it}$, $\simeq 1/2$ of that of Tr. A. One can see that the intensity for Tr. B is about 1/2 of that of Tr. A, indicating that the CP EDMR intensity is proportional to the $N_{it}$. In addition, the profiles of the spectra (black curve) for Tr. A and Tr. B are the same. After deconvolution of the profiles, we found that the $g$ value and the linewidth are all the same between Tr. A and Tr. B.

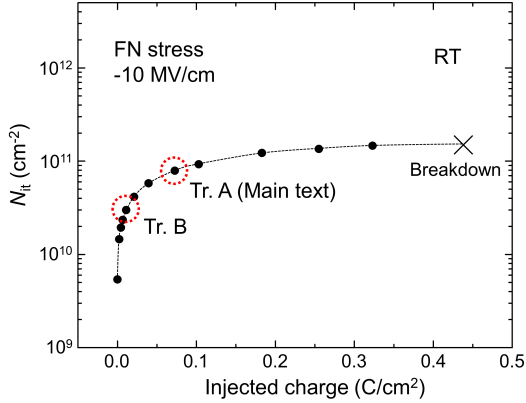

FIG. 8. Injected-charge dependence of defect density $N_{it}$. For the FN stress, the negatively-high voltage fixed at –30 V was applied to the gate, which corresponds to the electric field of –10 MV/cm applied to the oxide. $N_{it}$ is estimated from the CP method at room temperature. The red circles mark the $N_{it}$'s for Tr. A and Tr. B. The cross shows the breakdown point ($N_{it} \sim 1.5 \times 10^{11}$ cm$^{-2}$). The dashed line is a guide for the eye.

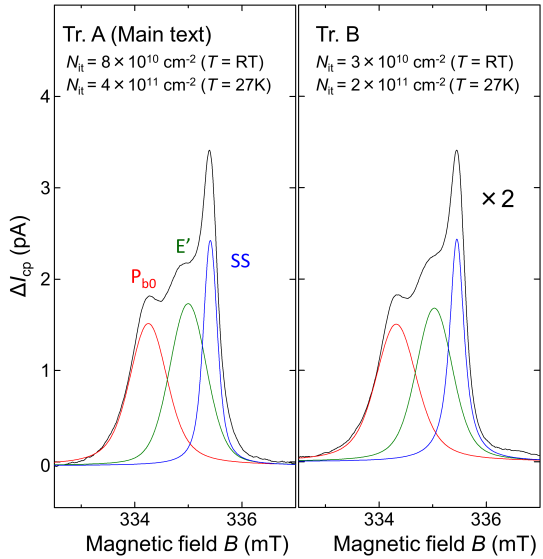

FIG. 9. CP EDMR characteristics of MOSFETs with different amounts of the injected charges. Integrated spectra for Tr. A (left panel) and Tr. B (right) measured at $T$ = 27 K with $\boldsymbol{B}$∥[100]. Parameters for the gate pulse are $f$ = 5 kHz, $\Delta V$ = 4 V, and $\Delta t$ = 30 µs. The black curves are the experimental data. Three peaks in each panel are decomposed using the Voigt function, each of which is assigned to the signal from $P_{b0}$ (red), E' (green) and SS (blue). The defect density $N_{it}$ for Tr. A is twice larger than that for Tr. B. The vertical axis for Tr. B is enlarged two times.

(ii) *Dependence on stress-bias polarity*

The third MOSFET, referred to as Tr. C, was stressed under the positive $V_g$ of +30 V for about 1 minute, and then under the negative $V_g$ of –15 V for a few minutes. The resultant area density $N_{it}$ at 27 K was estimated to be $3 \times 10^{11}$ cm$^{-2}$.

Figure 10 shows the CP EDMR spectra for Tr. A (left) and C (right) measured at 27 K. One can see that the profiles of the spectrum are different from each other. However, we found that, similarly to the case of Tr. A, three peaks gave us the best fit to the experimental data in Tr. C, and the g value of each peak agreed with that of $P_{b0}$ (red), E' (green) and SS (blue line) for the case of Tr. A. Note that the intensity (vertical axis) is normalized so that the peak height $\Delta I_{cp}$(SS) of the SS peak becomes the same for the two transistors. As one can see in the figure, the relative intensity between $P_{b0}$ and E' centers is different from each other between Tr. A and Tr. C. However, the ratio $R = \Delta I_{cp}$(SS)/($\Delta I_{cp}$($P_{b0}$) + $\Delta I_{cp}$(E')) was found to be kept constant at about 0.75. As described in the main text, this result supports the idea that SS-electrons can be the pair partners of both $P_{b0}$ and E' electrons.

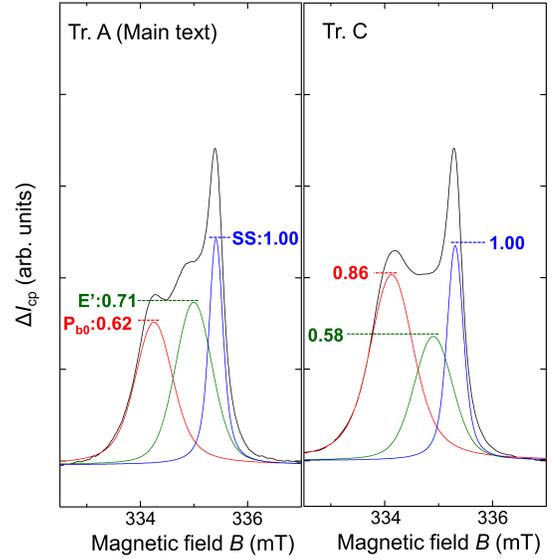

FIG. 10. CP EDMR characteristics of MOSFETs FN-stressed under different bias polarity. Integrated spectra for Tr. A (left panel) and Tr. C (right) measured at $T$ = 27 K with $\boldsymbol{B}$∥[100]. Parameters for the gate pulse are $f$ = 5 kHz, $\Delta V$ = 4 V, and $\Delta t$ = 30 µs. The black curves are the experimental data. Three peaks in each panel are decomposed using the Voigt function, each of which is assigned to the signal from $P_{b0}$ (red), E' (green) and SS (blue). Values shown in the figure are the relative intensities (relative peak heights) with that for the SS set at 1.00.

## REFERENCES

[1] P. Balk, The Si-SiO$_2$ system (ELSEVIER, Amsterdam, 1988).

[2] C. R. Helms and E. H. Poindexter, The silicon-silicon dioxide system: Its microstructure and imperfections. Rep. Prog. Phys. **57**, 791 (1994).

[3] G. D. Wilk, R. M. Wallace, and J. M. Anthony, High-k gate dielectrics: Current status and materials properties considerations. J. Appl. Phys. **89**, 5243 (2001).

[4] G. Ribes, J. Mitard, M. Denais, S. Bruyere, F. Monsieur, C. Parthasarathy, E. Vincent, and G. Ghibaudo, Review on high-k dielectrics reliability issues. IEEE Trans. Device Mater. **5**, 5 (2005).

[5] F. A. Zwanenburg, A. S. Dzurak, A. Morello, M. Y. Simmons, L. C. L. Hollenberg, G. Klimeck, S. Rogge, S. N. Coppersmith, and M. A. Eriksson, Silicon quantum electronics. Rev. Mod. Phys. **85**, 961 (2013).

[6] G. Yamahata, K. Nishiguchi, and A. Fujiwara, Gigahertz single-trap electron pumps in silicon. Nat. Commun. **5**, 6038 (2014).

[7] M. Xiao, I. Martin, E. Yablonovitch, and H. W. Jiang, Electrical detection of the spin resonance of a single electron in a silicon field-effect transistor. Nature **430**, 435 (2004).

[8] A. Dusko, A. L. Saraiva, and B. Koiller, Splitting valleys in Si/SiO$_2$: Identification and control of interface states. Phys. Rev. B **89**, 205307 (2014).

[9] J. S. Brugler and P. G. A. Jespers, Charge pumping in MOS devices. IEEE Trans. Electron Dev. **16**, 297 (1969).

[10] G. Groeseneken, H. E. Maes, N. Beltran, and R. F. DeKeersmaecker, A reliable approach to charge-pumping measurements in MOS transistors. IEEE Trans. Electron Dev. **31**, 42 (1984).

[11] N. S. Saks and M. G. Ancona, Determination of interface trap capture cross sections using three-level charge pumping. IEEE Electron Dev. Lett. **11**, 339 (1990).

[12] G. Van den Bosch, G. Groeseneken, P. Heremans, and H. E. Maes, Spectroscopic charge pumping: A new procedure for measuring interface trap distributions on MOS transistors. IEEE Trans. Electron Dev. **38**, 1820 (1991).

[13] N. S. Saks, G. Groeseneken, and I. DeWolf, Characterization of individual interface traps with charge pumping. Appl. Phys. Lett. **68**, 1383 (1996).

[14] N. S. Saks, Measurement of single interface trap capture cross sections with charge pumping. Appl. Phys. Lett. **70**, 3380 (1997).

[15] J. T. Ryan, L. C. Yu, J. H. Han, J. J. Kopanski, K. P. Cheung, F. Zhang, C. Wang, J. P. Campbell, and J. S. Suehle, Spectroscopic charge pumping investigation of the amphoteric nature of Si/SiO$_2$ interface states. Appl. Phys. Lett. **98**, 233502 (2011).

[16] M. Hori, T. Watanabe, T. Tsuchiya, and Y. Ono, Analysis of electron capture process in charge pumping sequence using time domain measurements. Appl. Phys. Lett. **105**, 261602 (2014).

[17] T. Tsuchiya and Y. Ono, Charge pumping current from single Si/SiO$_2$ interface traps: Direct observation of P$_b$ centers and fundamental trap-counting by the charge pumping method. Jpn. J. Appl. Phys. **54**, 04DC01 (2015).

[18] T. Tsuchiya and P. M. Lenahan, Distribution of the energy levels of individual interface traps and a fundamental refinement in charge pumping theory. Jpn. J. Appl. Phys. **56**, 031301 (2017).

[19] M. Xiao, I. Martin, and H. W. Jiang, Probing the spin state of a single electron trap by random telegraph signal. Phys. Rev. Lett. **91**, 078301 (2003).

[20] J. T. Ryan, P. M. Lenahan, A. T. Krishnan, and S. Krishnan, Spin dependent tunneling spectroscopy in 1.2 nm dielectrics. J. Appl. Phys. **108**, 064511 (2010).

[21] P. Broqvist, A. Alkauskas, and A. Pasquarello, Defect levels of dangling bonds in silicon and germanium through hybrid functionals. Phys. Rev. B **78**, 075203 (2008).

[22] Y. Nishi, Study of silicon-silicon dioxide structure by electron spin resonance I. Jpn. J. Appl. Phys. **10**, 52 (1971).

[23] M. Stutzmann, M. S. Brandt, and M. W. Bayerl, Spin-dependent processes in amorphous and microcrystalline silicon: a survey. J. Non-Cryst. Solids **266-269**, 1 (2000).

[24] J. H. Stathis and H. Nishikawa, discussed at the IEEE Semiconductor Interface Specialists Conference (SISC), San Diego, CA, 3 to 5 December 1998.

[25] B. C. Bittel, P. M. Lenahan, J. T. Ryan, J. Fronheiser, and A. Lelis, Spin dependent charge pumping in SiC metal-oxide-semiconductor field-effect-transistors. J. Appl. Phys. Lett. **99**, 083504 (2011).

[26] M. A. Anders, P. M. Lenahan, and A. J. Lelis, Multi-resonance frequency spin dependent charge pumping and spin dependent recombination - applied to the 4H-SiC/SiO$_2$ interface. J. Appl. Phys. **122**, 234503 (2017).

[27] M. Hori, T. Tsuchiya, and Y. Ono, Improvement of charge-pumping electrically detected magnetic resonance and its application to silicon metal-oxide-semiconductor field-effect transistor. Appl. Phys. Expr. **10**, 015701 (2017).

[28] E. H. Poindexter, P. J. Caplan, B. E. Deal, and R. R. Razouk, Interface states and electron spin resonance centers in thermally oxidized (111) and (100) silicon wafers. J. Appl. Phys. **52**, 879 (1981).

[29] P. M. Lenahan and J. F. Jr. Conley, What can electron paramagnetic resonance tell us about the Si/SiO$_2$ system? J. Vac. Sci. Tecnol. B **16**, 2134 (1998).

[30] M. Jivanescu, A. Stesmans, and V. V. Afanas'ev, Multifrequency ESR analysis of the E'$_\delta$ defect in a-SiO$_2$. Phys. Rev. B **83**, 094118 (2011).

[31] C. Boehme, Dissertation, Philipps Universität Marburg (2002).

[32] D. J. DiMaria and E. Cartier, Mechanism for stress-induced leakage currents in thin silicon dioxide films. J. Appl. Phys. **78**, 3883 (1995).

[33] J. T. Krick, P. M. Lenahan, and D. J. Dunn, Direct observation of interfacial point defects generated by channel hot hole injection in n-channel metal oxide silicon field effect transistors. Appl. Phys. Lett. **59**, 3437 (1991).

[34] K. L. Brower, Strain broadening of the dangling-bond resonance at the (111)Si-SiO$_2$ interface. Phys. Rev. B **33**, 4471 (1986).

[35] J. H. Stathis and E. Cartier, Atomic hydrogen reactions with P$_b$ centers at the (100) Si/SiO$_2$ interface. Phys. Rev. Lett. **72**, 2745 (1994).

[36] S. Shankar, A. M. Tyryshkin, S. Avasthi, and S. A. Lyon, Spin resonance of 2D electrons in a large-area silicon MOSFET. Physica E **40**, 1659 (2008).

[37] S. Shankar, A. M. Tyryshkin, J. He, and S. A. Lyon, Spin relaxation and coherence times for electrons at the Si/SiO$_2$ interface. Phys. Rev. B **82**, 195323 (2010).

[38] C. F. Young, E. H. Poindexter, and G. J. Gerardi, Electron paramagnetic resonance of porous silicon: Observation and identification of conduction-band electrons. J. Appl. Phys. **81**, 7468 (1997).

[39] R. E. Paulsen and M. H. White, Theory and application of charge pumping for the characterization of Si-SiO$_2$ interface and near-interface oxide traps. IEEE Trans. Electron Dev. **41**, 1213 (1994).

[40] J. L. Cantin, M. Schoisswohl, H. J. Bardeleben, N. H. Zoubir, and M. Vergnat, Electron-paramagnetic-resonance study of the microscopic structure of the Si(001)-SiO$_2$ interface. Phys. Rev. B **52**, R11599 (1995).

[41] K. P. O'Donnell and X. Chen, Temperature dependence of semiconductor band gaps. Appl. Phys. Lett. **58**, 2924 (1991).

[42] P. M. Lenahan, Atomic scale defects involved in MOS reliability problems. Microelectron. Eng. **69**, 173 (2003).

[43] D. J. Lepine, Spin-dependent recombination on silicon surface. Phys. Rev. B **6**, 436 (1972).

[44] D. Kaplan, I. Solomon, and N. F. Mott, Explanation of the large spin-dependent recombination effect in semiconductors. J. Phys. Lett. **39**, 51 (1977).

[45] H. Dersch, L. Schweitzer, and J. Stuke, Recombination processes in a-Si:H: Spin-dependent photoconductivity. Phys. Rev. B **28**, 4678 (1983).

[46] F. C. Rong, W. R. Buchwald, E. H. Poindexter, W. L. Warren, and D. J. Keeble, Spin-dependent Shockley-Read recombination of electrons and holes in indirect-band-gap semiconductor p-n junction diodes. Solid-State Electron. **34**, 835 (1991).

[47] I. Hiromitsu, Y. Kaimori, M. Kitano, and T. Ito, Spin-dependent recombination of photoinduced carriers in phthalocyanine/C$_{60}$ heterojunctions. Phys. Rev. B **59**, 2151 (1999).

[48] W. Fuhs, P. Kanschat, and K. Lips, Bandtails and defects in microcrystalline silicon (μc-Si:H). J. Vac. Sci. Technol. B **18**, 1792 (2000).

[49] K. Lips, P. Kanschat, and W. Fuhs, Defects and recombination in microcrystalline silicon. Sol. Eng. Mat. & Sol. Cells **78**, 513 (2003).

[50] C. Boehme and K. Lips, Theory of time-domain measurement of spin-dependent recombination with pulsed electrically detected magnetic resonance. Phys. Rev. B **68**, 245105 (2003).

[51] C. Boehme and K. Lips, A pulsed EDMR study of hydrogenated microcrystalline silicon at low temperatures. Phys. Stat. Sol. (c) **1**, 1255 (2004).

[52] F. Friedrich, C. Boehme, and K. Lips, Triplet recombination at P$_b$ centers and its implications for capture cross sections. J. Appl. Phys. **97**, 056101 (2005).

[53] A. R. Stegner, C. Boehme, H. Huebl, M. Stutzmann, K. Lips, and M. S. Brandt, Electrical detection of coherent $^{31}$P spin quantum states. Nat. Phys. **2**, 835 (2006).

[54] S. Y. Paik, S. Y. Lee, W. J. Baker, D. R. McCamey, and C. Boehme, $T_1$ and $T_2$ spin relaxation time limitations of phosphorous donor electrons near crystalline silicon to silicon dioxide interface defects. Phys. Rev. B **81**, 075214 (2010).

[55] M. Lenzlinger and E. H. Snow, Fowler-Nordheim tunneling into thermally grown SiO$_2$. J. Appl. Phys. **40**, 278 (1969).

[56] G. Krieger and R. M. Swanson, Fowler-Nordheim electron tunneling in thin Si-SiO$_2$-Al structures. J. Appl. Phys. **52**, 5710 (1981).

[57] D. Vuillaume, D. Goguenheim, and J. C. Bourgoin, Nature of the defects generated by electric field stress at the Si-SiO$_2$ interface. Appl. Phys. Lett. **58**, 490 (1991).

[58] D. J. DiMaria, Defect generation under substrate-hot-electron injection into ultrathin silicon dioxide layers. J. Appl. Phys. **86**, 2100 (1999).

[59] K. Kobayashi, A. Teramoto, and H. Miyoshi, Origin of positive charge generated in thin SiO$_2$ films during high-field electrical stress. IEEE Trans. Electron Dev. **46**, 947 (1999).